\documentclass{emulateapj}

\begin{document}

\shortauthors{Luhman et al.}
\shorttitle{Dynamical Decay in the Kleinmann-Low Nebula}

\title{New Evidence for the Dynamical Decay of a Multiple System in the Orion
Kleinmann-Low Nebula\altaffilmark{1}}

\author{
K. L. Luhman\altaffilmark{2,3},
M. Robberto\altaffilmark{4,5},
J. C. Tan\altaffilmark{6,7},
M. Andersen\altaffilmark{8},
M. Giulia Ubeira Gabellini\altaffilmark{4,9},
C. F. Manara\altaffilmark{10},
I. Platais\altaffilmark{11},
\& L. Ubeda\altaffilmark{4}}

\altaffiltext{1}
{Based on observations made with the NASA/ESA {\it Hubble Space Telescope}
and the NASA Infrared Telescope Facility.}

\altaffiltext{2}{Department of Astronomy and Astrophysics,
The Pennsylvania State University, University Park, PA 16802, USA;
kluhman@astro.psu.edu}
\altaffiltext{3}{Center for Exoplanets and Habitable Worlds, The
Pennsylvania State University, University Park, PA 16802, USA}
\altaffiltext{4}{Space Telescope Science Institute, 3700 San Martin Drive
Baltimore, MD 21218, USA}
\altaffiltext{5}{Johns Hopkins University, Center for Astrophysical Sciences
3400 North Charles Street, Baltimore, MD 21218, USA}
\altaffiltext{6}{Department of Astronomy, University of Florida, Gainesville,
FL 32611, USA}
\altaffiltext{7}{Department of Physics, University of Florida, Gainesville,
FL 32611, USA}
\altaffiltext{8}{Gemini Observatory, Casilla 603, La Serena, Chile}
\altaffiltext{9}{Dipartimento di Fisica, Universit\`a degli Studi di Milano,
via Celoria 16, I-20133 Milano, Italy}
\altaffiltext{10}{Scientific Support Office, Directorate of Science, European
Space Research and Technology Centre, Keplerlaan 1, 2201 AZ, Noordwijk,
The Netherlands}
\altaffiltext{11}{Johns Hopkins University, Department of Physics and Astronomy,
3400 North Charles Street, Baltimore, MD 21218, USA}

\begin{abstract}

We have measured astrometry for members of the Orion Nebula Cluster
with images obtained in 2015 with the Wide Field Camera 3 on board
the {\it Hubble Space Telescope}. By comparing those data to previous
measurements with NICMOS on {\it Hubble} in 1998, we have discovered that
a star in the Kleinmann-Low Nebula, source x from \citet{lon82}, is moving
with an unusually high proper motion of 29~mas~yr$^{-1}$, which corresponds
to 55~km~s$^{-1}$ at the distance of Orion.
Previous radio observations have found that three other stars in the 
Kleinmann-Low Nebula (BN and sources I and n) have high proper motions
(5--14~mas~yr$^{-1}$) and were near a single location $\sim$540 years ago,
and thus may have been members of a multiple system that dynamically decayed.
The proper motion of source x is consistent with ejection from that
same location 540 years ago, which provides strong evidence that
the dynamical decay did occur and that the runaway star BN originated in the
Kleinmann-Low Nebula rather than the nearby Trapezium cluster.
However, our constraint on the motion of source~n
is significantly smaller than the most recent radio measurement,
which indicates that it did not participate in the event that ejected the other three stars.

\end{abstract}

\keywords{
astrometry ---
stars: formation ---
stars: kinematics and dynamics ---
stars: pre-main sequence ---
stars: protostars}

\section{Introduction}
\label{sec:intro}

The Kleinmann-Low (KL) Nebula is a source of extended infrared (IR) emission
that is located $1\arcmin$ northwest of the Trapezium stars in the Orion Nebula
Cluster \citep{kle67}.
The nebula is heavily embedded and contains a dense group of young stellar
objects, some of which appear to have masses of $\gtrsim10$~$M_\odot$,
making it the nearest site of massive star formation
\citep[$\sim400$~pc,][]{men07,kim08,kou17}.
For one of these massive stars, the Becklin-Neugebauer (BN) object
\citep{bec67}, multi-epoch radio observations have revealed an unusually
high proper motion \citep{pla95}.
Members of Orion exhibit a dispersion of $\sim1$~mas~yr$^{-1}$ in
their proper motions \citep{van88,dzi17}, whereas BN has a proper motion of
13.5~mas~yr$^{-1}$ (26~km~s$^{-1}$) relative to the cluster \citep{rod17}.

Runaway young stars may arise from dynamical interactions within
multiple systems or small stellar groups \citep{pov67}.
Because its proper motion is pointed away from the Trapezium stars,
it has been proposed that BN was ejected 4000 years ago from
a multiple system containing the most massive member of the Trapezium,
$\theta^1$~Ori~C \citep{pla95,tan04}, which also helps explain
some of the properties of that system \citep{cha12}.
The proper motion of BN indicates that it may have passed near
radio source I \citep{gar87,chu87} in the KL Nebula $\sim$500 years ago.
\citet{tan04} suggested that following its ejection from the
Trapezium, BN passed close enough to source I to trigger the explosive
outflow that emanates from the cloud core in the KL nebula \citep{all93}.
Kinematic studies of the outflow have found that it did originate in
the vicinity of source~I near that time period \citep{zap09,bal11}.
Alternatively, \citet{bal05} and \citet{rod05} proposed that BN, source I,
and at least one additional star were members of a multiple system, and the
latter two stars formed a tight binary or merged, which resulted in the
ejection of BN and the generation of the outflow. 
That scenario has been supported by the measurement of source I's proper
motion, which is in roughly the opposite direction of BN's motion
\citep{rod05,god11}.
Based on radio measurements of its proper motion, source n from \citet{lon82}
also may have originated from a multiple system with source I and BN
\citep{gom05,gom08,rod17}, although its motion is more difficult to reliably
measure because of its complex and evolving morphology at radio wavelengths
\citep{god11}.

Previous studies have generally favored the KL Nebula rather
than the Trapezium as the ejection site for BN \citep{bal05,gom08,god11}.
For instance, BN appears to be younger than the Trapezium stars based 
on its circumstellar material and its ejection from the Trapezium and
subsequent close encounter with source I would not account for the large
motions of I and n. 
Nevertheless, some doubts persist regarding this scenario in
which BN was ejected from a multiple system with source I \citep{god11,pla17}.
It is challenging to explain the presence of the circumstellar
disks around BN and sources I and n if they experienced interactions that
were sufficiently close to produce the rapid ejection of BN. In addition,
the mass of source I derived by combining an estimate of BN's mass with
conservation of momentum ($\sim20$~$M_\odot$) is higher than the dynamical
mass measured from its circumstellar material
\citep[$\sim7$~$M_\odot$,][]{mat10,hir14,pla17}.

The proper motions of BN, source I, and other stars embedded in the
KL Nebula have been previously measured with multi-epoch
radio observations. However, many young stars in the nebula
are not detected in radio surveys, and instead appear only in IR imaging.
To search for large motions among those stars, we have used
high-resolution near-IR images that were obtained with the {\it Hubble Space 
Telescope} ({\it HST}) in 1998 and 2015. In this Letter, we report the
discovery of a high proper motion for source x from \citet{lon82}.
Based on its motion, source x likely originated from the same multiple
system that produced BN and source I. Meanwhile, we find that the proper motion of
source~n is inconsistent with participation in that event.

\section{Astrometry in the Kleinmann-Low Nebula}

To search for high proper motion stars in the KL Nebula that are not
detected by radio surveys, we have used images spanning 17 years that were
obtained with the Near-Infrared Camera and Multi-Object Spectrometer (NICMOS)
and the Wide Field Camera 3 \citep[WFC3,][]{kimb08} on board {\it HST}.
On 1998 January 12 and 16 (UT), camera 3 of NICMOS collected images of the KL Nebula in the
filters F110W, F160W, F164N, F166N, F190N, F212N, and F215N \citep{sch99,luh00}.
On 2015 March 13 (UT), WFC3 observed a much larger portion of the
Orion Nebula Cluster, including the entirety of the KL Nebula,
in the F130N and F139M filters through Treasury program 13826 (M. Robberto,
in preparation).
The raw images from NICMOS and WFC3 had plate scales of approximately
$0.2\arcsec$~pixel$^{-1}$ and $0.13\arcsec$~pixel$^{-1}$, respectively.

Pixel coordinates for point sources in each of the NICMOS
and WFC3 images were measured with the task {\it starfind} in IRAF.
We used astrometry from the near-IR imaging survey of \citet{mei16} for stars 
in our F130N and F139M WFC3 images to derive offsets in right ascension,
declination, and rotation that would align the World Coordinate System (WCS)
of the WFC3 images to the astrometric system utilized by \citet{mei16}, which
was that of the Two Micron All-Sky Survey \citep{skr06}. We measured the WCS of
each NICMOS image using astrometry of stars derived from those updated WFC3 images.

For each of the two cameras,
we computed the average
coordinates of each star from among the filters in which it was detected, excluding
data in the corners of the NICMOS images because of the greater uncertainty in the
distortion correction.
For each star detected by WFC3, we then identified the closest matching star
in the catalog of sources from NICMOS. Proper motions were computed with
the matched WFC3 and NICMOS coordinates.
We estimated the errors in these motions from the standard deviations 
of the differences in right ascension and declination between the two epochs.
These estimates are upper limits for the errors since the velocity
dispersion among the stars \citep[$\sim1$~mas~yr$^{-1}$,][]{van88,dzi17}
also contributes to the dispersion in our proper motion measurements.

Among the stars in the vicinity of the KL Nebula, source x from \citet{lon82}
exhibits the largest motion.
Its total proper motion is 29~mas~yr$^{-1}$, which corresponds to a velocity
of 55~km~s$^{-1}$ at the distance of Orion. Foreground field stars often
have large proper motions, but source x cannot be a foreground star given
the high extinction indicated by its red colors (Section~\ref{sec:phot}).
Its IR excess emission and spectral features provide additional evidence of youth
and membership in Orion (Sections~\ref{sec:phot} and \ref{sec:spec}).
In Table~\ref{tab:data}, we list the position of source~x in the
WFC3 data and its proper motion in right ascension and declination.

BN has the second largest proper motion among stars in the KL Nebula that
were detected by WFC3 and NICMOS. We measure a proper motion of
($\mu_\alpha$,$\mu_\delta$)=($-6.9\pm1.4$,$+10.3\pm1.4$~mas~yr$^{-1}$) for BN
in the rest frame of Orion, which is consistent with the more accurate
measurement at radio wavelengths from \citet{rod17}.
We did not find any additional stars in the KL Nebula with relative proper
motions larger than $2\sigma$ ($\gtrsim3$~mas~yr$^{-1}$).

As discussed in Section~\ref{sec:intro}, some previous studies have reported a fairly
large proper motion for source n \citep[$0\pm0.9$,$-7.8\pm0.6$~mas~yr$^{-1}$,][]{rod17}.
However, we derive a smaller value of $-1.8\pm1.4$,$-2.5\pm1.4$~mas~yr$^{-1}$,
which agrees better with the motion of $+1.6\pm1.6$,$+3.4\pm1.6$~mas~yr$^{-1}$ 
measured from millimeter data by \citet{god11}.
In centimeter continuum images, source~n exhibited two components separated by
$\sim0\farcs35$ in early epochs \citep{gom05} and appeared as a single elongated object
in later data \citep{gom08,rod17}. It is unclear which, if any, of the flux peaks
in those images correspond to the star, which makes it difficult to reliably measure
the star's motion with those data. \citet{rod17} refer to source~n as a double star based
on the radio morphology, but it is a single point source in the {\it HST} images.

\begin{figure}
\epsscale{1}
\plotone{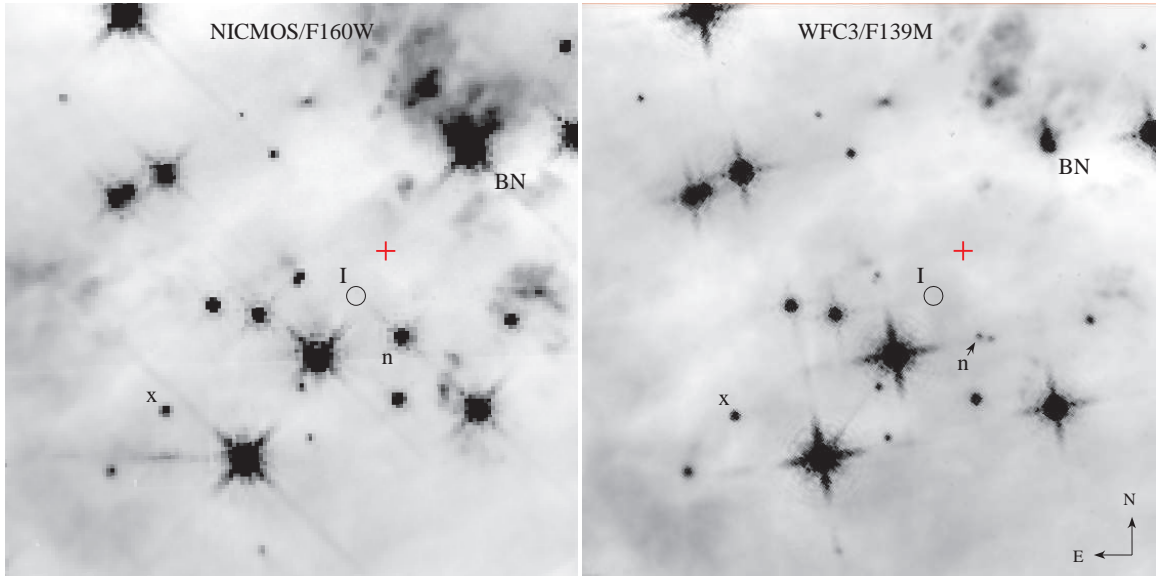}
\caption{
Near-IR images of a portion of the KL Nebula from NICMOS in 1998
and WFC3 in 2015. BN and sources n and x are indicated.
Source I was not detected in these data;
its position at radio wavelengths is circled \citep{rod17}. 
We have marked the initial position of BN and source I in 1475 that
was proposed by \citet[][cross]{rod17}.
The size of each image is $30\arcsec\times30\arcsec$.
}
\label{fig:images}
\end{figure}

In Figure~\ref{fig:images}, we show the NICMOS/F160W and WFC3/F139M images
for a $30\arcsec\times30\arcsec$ field encompassing BN and sources I, n, and x.
Source I was not detected in these data (or any previous IR images).

\section{Photometry of Source x}
\label{sec:phot}

Since its initial detection at 1.6 and 2.2~\micron\ by \citet{lon82},
source x has been imaged in X-rays \citep{get05} and in bands from
1.6--11.7~\micron\ \citep{hil00,luh00,lad00,lad04,mue02,rob05,rob10,smi05,mei16}.
It has exhibited variations of $\sim1$~mag among the multiple measurements
that are available in the $H$ and $K$ bands. As a result, we have adopted
the median of the available $H-K$ and $H-K_s$ colors in which the bands
were measured near the same time, and we have adopted the median of the
available $K$ and $K_s$ data.
In Table~\ref{tab:data}, we have compiled those median values and
the (single) measurements of $m_{160}$, $K-L$, $N$, and $F_{11.7}$. 
Source x was not detected in the F110W image from NICMOS.
We note that \citet{fav11} detected methyl formate emission at $1\farcs9$ from 
source~x. Two other stars also appear within 3--$4\arcsec$ of that emission peak,
so it is unclear whether it is associated with source~x.

We have measured aperture photometry for source x in the WFC3 images
using an aperture radius of two pixels and radii of two and
five pixels for the inner and outer boundaries of the sky annulus, respectively.
We estimated aperture corrections of 0.166 (F130N) and 0.176 (F139M)
between those apertures and radii of $0\farcs4$ using bright stars in the
WFC3 images. We applied those corrections and the zero-point Vega magnitudes
of 21.8258 (F130N) and 23.2093 (F139M) for $0\farcs4$
apertures\footnote{http://www.stsci.edu/hst/wfc3/phot\_zp\_lbn}
to the photometry of source x. The resulting measurements of $m_{130}$
and $m_{139}$ are included in Table~\ref{tab:data}.

\begin{figure}
\epsscale{1}
\plotone{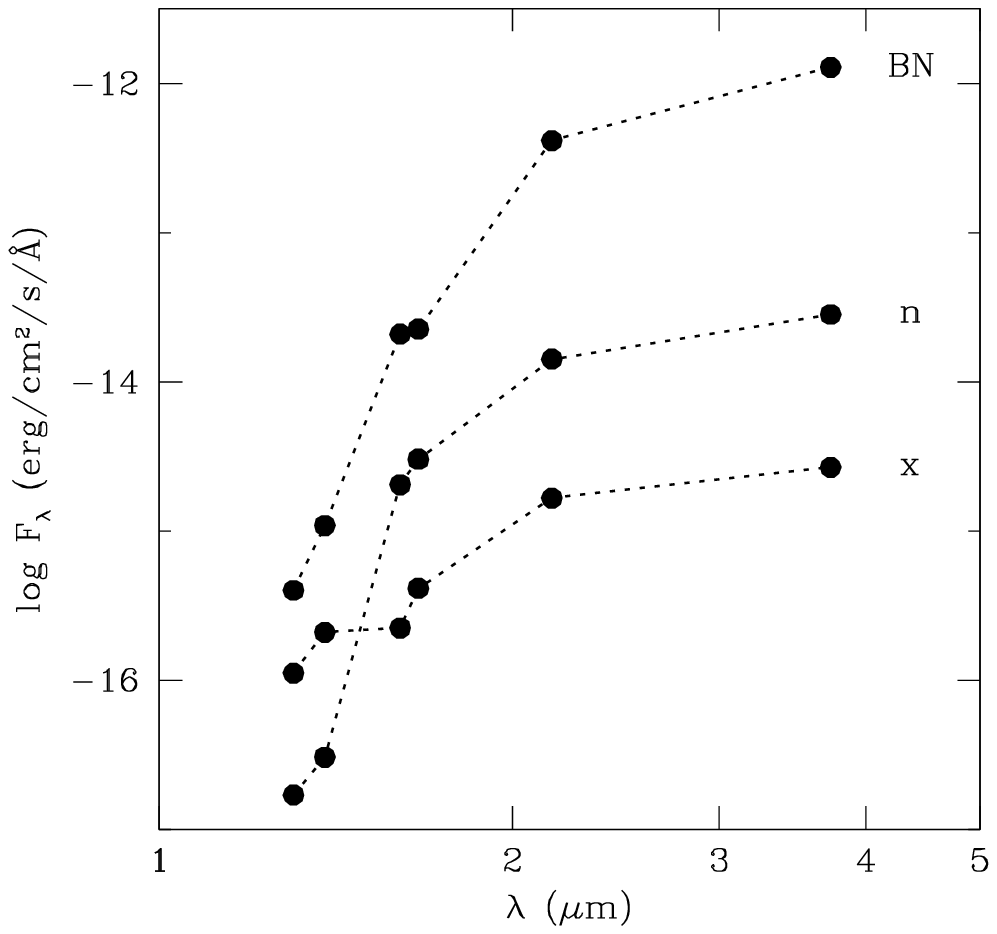}
\caption{
Near-IR SEDs for BN and sources n and x
\citep[Table~\ref{tab:data}, 2MASS,][]{sim99,hil00,lad00,luh00,mue02,rob05}.
The kink in the SED of source~x at shorter wavelengths may indicate that it became brighter
between the older observations at $\lambda\geq1.6$~\micron\ and the more
recent images in F130N and F139M with WFC3.
}
\label{fig:sed}
\end{figure}

To compare the evolutionary stage of source x to those of BN and source n, we
have plotted their near-IR spectral energy distributions (SEDs) in
Figure~\ref{fig:sed}.
When using bands that are measured near the same time, the colors of
source x are similar to or slightly bluer than those of BN and source n,
indicating that source x resembles the latter stars in having high
extinction and IR excess emission from circumstellar material.
For instance, \citet{lad00} classified source x as a candidate protostar
based on its red $K-L$ color.
However, source x is much bluer than BN and source n between the WFC3 filters
and bands at longer wavelengths, as illustrated in Figure~\ref{fig:sed},
which indicates that source x probably brightened by a few magnitudes
prior to the WFC3 observations.

\section{Spectroscopy of Source x}
\label{sec:spec}

Since spectroscopy has not been previously reported for source~x,
we observed it with the near-IR spectrograph 
SpeX \citep{ray03} at the NASA Infrared Telescope Facility (IRTF)
on the night of 2017 January 14. The instrument was operated in
the SXD mode with the $0\farcs8$ slit, which produced data with a resolution
of $R\sim800$ and a wavelength coverage of 0.8--2.5~\micron.
We collected ten 1~min exposures in an ABBA pattern along the slit.
To facilitate removal of nebular line emission 
at the location of source~x, we selected A/B positions on the slit that were
fairly close to each other ($3\arcsec$) and a position angle for the slit ($30\arcdeg$)
that minimized contamination from extended H$_2$ emission \citep{sto98,sch99}.

The data for source~x were reduced with the Spextool package \citep{cus04} and
corrected for telluric absorption in the manner described by \citet{vac03}.
The spectral images in the A and
B slit positions were subtracted from each other to remove sky emission. Based on the
spectra at positions adjacent to source~x along the slit, the extended hydrogen
and helium emission from the Orion Nebula was successfully removed, and the remaining
emission in the spectrum of source~x should arise from the star. However, because the H$_2$
emission across the KL Nebula varies on small angular scales, our method
of sky subtraction produced erroneous negative residuals in the spectrum of source~x at the
wavelengths of H$_2$ lines. Therefore, we have ignored the data at those wavelengths.

\begin{figure}
\epsscale{1}
\plotone{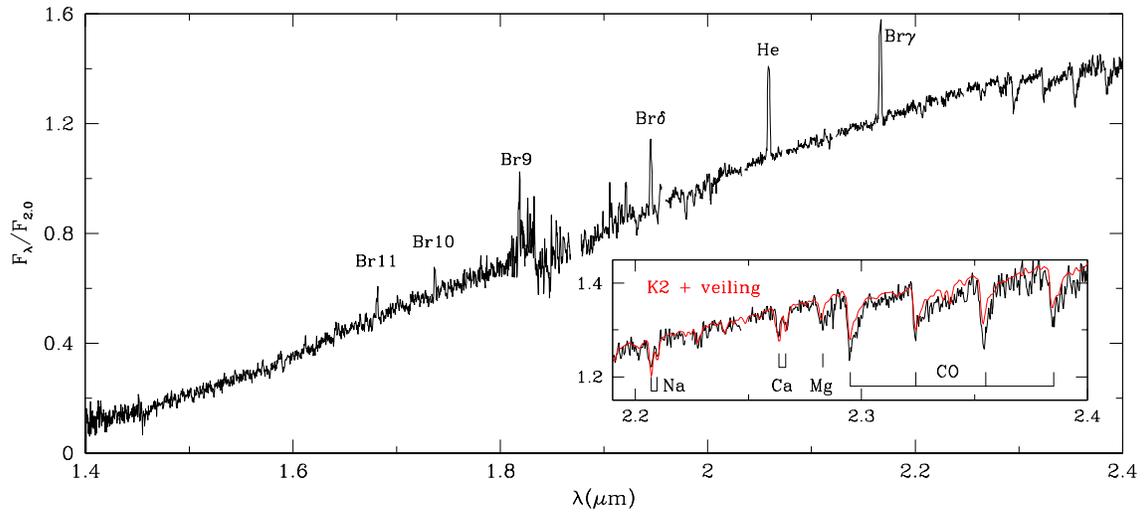}
\caption{
Near-IR spectrum of source x ($R\sim800$). The absorption features are
indicative of an early K star with continuum veiling, as illustrated in the
inset, where we have applied veiling to a spectrum of
LkCa~19 \citep[K2,][]{her14} from M. McClure (private communication) to match
the strengths of Na, Ca, and Mg in source~x.
The data used to create this figure are available.
}
\label{fig:spec}
\end{figure}

The reduced spectrum of source~x from 1.4--2.4~\micron\ is shown in Figure~\ref{fig:spec}.
Little flux was detected at shorter wavelengths.
In addition to the hydrogen and helium emission lines mentioned earlier, the spectrum
exhibits several absorption features from metals and CO.
Given the presence of IR excess emission in its SED, these features may be diluted
by continuum emission from circumstellar dust.
To estimate the spectral type in a way that is independent of such veiling, we
have compared the relative strengths of Na~I, Ca~I, and Mg~I at 2.2--2.3~\micron\ to
those measured from SpeX data for field dwarfs \citep{ray09} and diskless young
stars (M. McClure, private communication).
We arrive at a spectral type of early K for source~x. Its spectrum contains continuum
veiling according to both samples of standard stars, but the young stars
have stronger atomic lines than the dwarfs, so they imply greater veiling.
To illustrate one of the better fits,
we have included in Figure~\ref{fig:spec} a comparison of source~x and the young star
LkCa~19 \citep[K2,][]{her14}. The latter has been artificially veiled to match the
strengths of Na~I, Ca~I, and Mg~I in source~x ($F_{excess}/F_*=0.6$).
The ratio of the CO bandhead to the atomic lines is higher in source~x than in
the standards, which indicates a lower surface gravity.
The spectral slope of source~x indicates an extinction of $A_V\sim30$ based on
comparison to the best-fitting veiled standards.

\section{Kinematics of BN and Sources I and x}

\begin{figure}
\epsscale{1}
\plotone{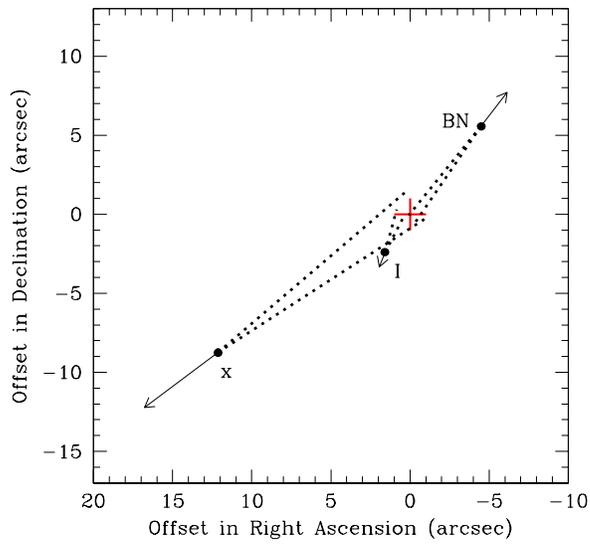}
\caption{
Positions of BN and sources I and x (filled circles) for
the epoch of the WFC3 images (2015.20) based on astrometry from
this work and \citet{rod17}. 
The positions are plotted relative to the initial position of BN and I
in 1475 that was proposed by \citet[][cross]{rod17}.
We also show the proper motions in the rest frame of Orion for
a period of 200 years (solid lines and arrows) and the range of allowed
paths (1~$\sigma$) back to 1475 based on those proper motion measurements
(dotted lines). The motion of source x is consistent with an origin
at the same position and time as the other two stars.
}
\label{fig:map}
\end{figure}

In Figure~\ref{fig:map}, we plot the positions and proper motion vectors of
BN and sources I and x.
We have excluded source~n since its proper motion constraints from the {\it HST}
images and the millimeter data from \citet{god11} are significantly
smaller than the other measurements that had suggested an origin at the same
location as BN and source~I \citep{gom05,gom08,rod17}.
Based on their relative proper motions, BN and source I experienced their
closest approach in the year 1475$\pm$6 with a projected separation of
$\lesssim$40~AU \citep[1~$\sigma$,][]{rod17}.
We have included in Figure~\ref{fig:map} the 1~$\sigma$ range of allowed
paths for each star back to that year.
The estimated location of source~x in 1475 agrees with the initial position for
the other two stars, which indicates that it was ejected in the same event as
those stars. Source x provides strong evidence that such an event did occur
and that BN originated in the KL Nebula rather than the Trapezium.

If BN and sources I and x were ejected from a multiple system,
they should exhibit little net momentum in the rest frame of Orion.
A radial velocity measurement is not available for source~x, but we can
estimate the net momentum in the plane of the sky with the proper motions.
We adopt 10~$M_\odot$ for BN based on its luminosity and ionization rate
\citep{rod05} and 7~$M_\odot$ for source~I based on the kinematics of its
circumstellar material \citep{mat10,hir14,pla17}.  We derive a luminosity
of $\sim20$~$L_\odot$ for source~x using its $H$-band magnitude, a bolometric
correction for an early K star, and the extinction from our spectrum.
By combining that luminosity and the temperature corresponding to an early K
star \citep[$\sim5000$~K,][]{sk82} with theoretical evolutionary models 
\citep{pal99}, we estimate a mass of 2.5--3~$M_\odot$ for source~x.
Using these adopted masses and the proper motions, the three stars exhibit
a momentum equivalent to a velocity of $\sim1.4$~km~s$^{-1}$ for the original 
system in the rest frame of Orion, which is comparable to the velocity
dispersion of the cluster. By accounting for the kinematics of source x,
we have resolved the previous discrepancy between the dynamical mass 
of 7~$M_\odot$ for source I and the mass of $\sim20$~$M_\odot$ implied
by conservation of momentum with BN alone \citep{gom08,god11}.

Some aspects of the interaction that ejected BN and sources I and x remain
unclear, such as how these stars were able to retain or reform their
circumstellar disks \citep{god11,pla17} and the source of their current
kinetic energy.
As discussed in Section~\ref{sec:intro}, the latter has been previously
attributed to the gravitational potential energy released by a pair of stars
forming a tight binary or merging, likely in what is now source I.
The energy produced by the merger of stars with
a total mass of source~I (e.g., 1 and 6~$M_\odot$) is an order of magnitude
greater than the kinetic energy of the three stars, so that remains a
plausible explanation.

\acknowledgements
This work was supported by grant AST-1208239 from the NSF
and grant GO-13826 from the Space Telescope Science Institute.
The NASA/ESA {\it HST} is operated by the Space Telescope
Science Institute and the Association of Universities for Research in
Astronomy, Inc., under NASA contract NAS 5-26555.
The IRTF is operated by the University of Hawaii under contract NNH14CK55B
with NASA.
The Center for Exoplanets and Habitable Worlds is supported by the
Pennsylvania State University, the Eberly College of Science, and the
Pennsylvania Space Grant Consortium.

\clearpage

\begin{deluxetable}{lll}
\tabletypesize{\scriptsize}
\tablewidth{0pt}
\tablecaption{Astrometry and Photometry of Source x\label{tab:data}}
\tablehead{
\colhead{Parameter} & \colhead{Value} & \colhead{Reference}}
\startdata
$\alpha$(J2000, epoch 2015.20) & 5$^{\rm h}$35$^{\rm m}$15.222$^{\rm s}$ & 1 \\
$\delta$(J2000, epoch 2015.20) & $-5\arcdeg$22$\arcmin$36.96$\arcsec$ & 1 \\
$\mu_{\alpha}$ cos $\delta$\tablenotemark{a} & $+23.3\pm1.4$~mas~yr$^{-1}$ & 1 \\
$\mu_{\delta}$\tablenotemark{a} & $-17.4\pm$1.4~mas~yr$^{-1}$ & 1 \\
$m_{130}$ & 15.93$\pm$0.02 mag & 1 \\
$m_{139}$ & 15.01$\pm$0.02 mag & 1 \\
$m_{160}$ & 14.38$\pm$0.02 mag & 2 \\
$H-K$ and $H-K_s$(median) & 2.58 mag & 3,4,5 \\
$K$ and $K_s$(median) & 11.04 mag & 3,4,5,6 \\
$K-L$ & 2.76$\pm0.06$ mag & 4,7\\
$N$ & 6.59$\pm$0.1 mag & 8 \\
$F_{\nu}$(11.7$\micron$) & 0.16 Jy & 9 
\enddata
\tablenotetext{a}{Proper motion in the rest frame of Orion.}
\tablerefs{
(1) this work;
(2) \citet{luh00};
(3) \citet{hil00};
(4) \citet{mue02};
(5) \citet{rob10};
(6) \citet{sim99};
(7) \citet{lad00};
(8) \citet{rob05};
(9) \citet{smi05}.}
\end{deluxetable}

\end{document}